\documentclass{iopart}[twocolumn]

\usepackage{hyperref}
\usepackage[english]{babel}
\usepackage{graphicx}
\usepackage{color,xcolor}
\usepackage{siunitx}
\usepackage{csquotes}
\usepackage{subfig}
\usepackage{amssymb, amsfonts, hyperref, times}
\usepackage{bm}
\usepackage{tablefootnote}
\providecommand{\vect}[1]{{\boldsymbol{#1}}}

\begin{document}

\title{Current-induced H-shaped-skyrmion creation and their dynamics in the helical phase }

\author{R.~Knapman$^1$, D.~R.~Rodrigues$^1$, J.~Masell$^2$, and K.~Everschor-Sitte$^3$}

\address{$^1$ Institute of Physics, Johannes Gutenberg University Mainz, 55128 Mainz, Germany}
\address{$^2$ RIKEN Center for Emergent Matter Science (CEMS), Wako 351-0198, Japan}
\address{$^3$ Faculty of Physics, University of Duisburg-Essen, 47057 Duisburg, Germany}

\ead{knapman@uni-mainz.de}

\begin{abstract}
Inevitable for the basic principles of skyrmion racetrack-like applications is not only their confined motion along one-dimensional channels but also their controlled creation and annihilation. Helical magnets have been suggested to naturally confine the motion of skyrmions along the tracks formed by the helices, which also allow for high-speed skyrmion motion. We propose a protocol to create topological magnetic structures in a helical background.  We furthermore analyse the stability and current-driven motion of the skyrmions in a helical background with in-plane uniaxial anisotropy fixing the orientation of the helices.
\end{abstract}


\maketitle

%%%%%%%%%%%%%%%%%%%%%%
\section{Introduction}
%%%%%%%%%%%%%%%%%%%%%%

Magnetic skyrmions are topologically nontrivial textures that arise in magnetic materials.~\cite{Nagaosa2013, Fert2017, Everschor-Sitte2018, Back2020} They are typically considered to be vortex-like solitons in a ferromagnetic background. Owing to their small size and high stability against thermal fluctuations, they are considered as possible candidates for information units with applications in devices such as logic gates~\cite{Zhang2015l, Luo2018} and racetrack memory devices.~\cite{Fert2013, Tomasello2014,Lee2021}
The latter have gained considerable attention as candidates for low-input non-volatile memory. A major challenge for skyrmion-based racetracks is that the magnetic whirls typically not only move along the tracks but additionally experience current-induced forces in a perpendicular direction.~\cite{Jonietz2010, Everschor2011} This skyrmion Hall effect~\cite{Litzius2017, Jiang2017}
might lead to the annihilation of a skyrmion as it moves towards the edges of the racetrack. Several studies have been proposed to suppress or eliminate the skyrmion Hall effect which basically can be grouped into two categories. One exploits the coupling between two skyrmions of opposite polarity, as they arise for example in (artificial) antiferromagnets~\cite{Zhang2015b, Barker2016, Zhang2015j}, such that the Magnus forces acting on the individual skyrmions effectively cancel each other. The other one exploits an additional symmetry breaking which eliminates the skyrmion Hall effect along certain directions or by engineered material properties.~\cite{Hoffmann2017, Huang2017a, Kim2018, Gobel2019a, Zarzuela2020, Zhang2020d}
Another approach to avoid the annihilation of the skyrmion at the edge of the nanowire is to engineer the potential at the edges. The latter creates a repulsive force that counteracts the current-induced perpendicular force on the skyrmion.~\cite{Zhang2015c,Purnama2015, Muller2016,Lai2017,Tomasello2018a,Fook2015,Song2020,Gobel2021}.

The essential goal for racetrack memories is to devise a mechanism to constrain skyrmions to certain well-defined lanes.
While all of the suggested methods above counteract the Magnus force in some form, they often require fine-tuned material engineering and restrict the motion of the skyrmions along tracks only by means of geometrical confinement of the nanowire.
A different suggestion to obtain a skyrmion transport along 1D channels, which so far has not received much attention, is to exploit the tracks that are naturally formed in helical states.~\cite{Ezawa2011, Muller2017,Stepanova2021}.
In particular, it has been proposed that currents perpendicular to these formed tracks allow for high-speed skyrmion motion in materials with low damping. Furthermore, skyrmions and anti-skyrmions can coexist in helical magnets, and move along different tracks in the helical background. Any application, however, first requires the controlled creation of the topological objects.

In this work, we propose a protocol to create skyrmions in the helical background by electrical means. We show by numerical simulations that skyrmion-antiskyrmion pairs are nucleated at impurities, exploiting a concept that previously has been applied for ferromagnetic states.~\cite{Sitte2016,Muller2016a, Everschor-Sitte2016,Stier2017,Gao2019}. The generated topological structures immediately enter \enquote{lanes} that act as a natural confining potential along which the (anti-)skyrmions can be driven.

The manuscript is organized as follows. In Sec.~\ref{sec:model}, we consider the helical state in a thin film helimagnet with an in-plane easy axis. We discuss the helical ground state and different skyrmion solutions on top of the helical ground state. 
In Sec.~\ref{sec:dynamics} we analyse the current-induced dynamics of helical states and derive current-driven instabilities. Furthermore, we discuss the current-driven motion of skyrmions in the helical background. In Sec.~\ref{sec:shedding} we present an all-electrical protocol for creating skyrmions in the helical background.
The manuscript ends with a discussion and conclusion in Sec.~\ref{sec:disscon}.

%%%%%%%%%%%%%%%%%%%%%%%%%%%%%%%%%%%%%%%%%%%%%%%%%%%%%%%%%%%%%%%%%%%%%
\section{Topological excitations in helical states} \label{sec:model}
%%%%%%%%%%%%%%%%%%%%%%%%%%%%%%%%%%%%%%%%%%%%%%%%%%%%%%%%%%%%%%%%%%%%%

We consider a model for a smooth thin film helimagnet with an in-plane easy axis. The energy functional for this system is given by
\begin{equation}
\label{eq:Energy}
\fl E[\vect{m}(\vect{r})] = \int \rmd^2 r \, \left[ \frac{J}{2} (\nabla \vect{m})^2 + D \vect m \cdot \left(\hat{\vect{y}} \times \partial_{x}\vect{m} - \hat{\vect{x}} \times \partial_{y}\vect{m}\right) -K(\vect{m}\cdot\hat{\vect{y}})^2 \right],
\end{equation}
where  $\vect m(\vect{r})$ characterizes the normalized magnetization at position $\vect r=(x,y)$. The first term corresponds to the exchange interaction with strength $J$.  The second term is the chiral interaction corresponding to an interfacial Dzyaloshinskii-Moriya interaction (DMI) with strength $D$. The third term is the easy axis anisotropy along the in-plane direction $\hat{\vect{y}}$.
By choosing appropriate units Eq.~(\ref{eq:Energy}) can be transformed to a single-parameter theory. For this, we 
reparametrize energy and length by $J$ and $J/D$, respectively, to obtain 
\begin{eqnarray}
\label{eq:EnergyRep}
\fl E[\vect{m}(\vect{r})] = \int \rmd^2 r \, \left[ \frac{1}{2}\left(\partial_{x} \vect{m} - \hat{\vect{y}}\times \vect m \right)^2 + \frac{1}{2}\left(\partial_{y} \vect{m} + \hat{\vect{x}}\times \vect{m} \right)^2 + \frac{1}{2}m_{x}^2 \right. \\
    \left.  + \left(\frac{1}{2} - \kappa\right) m_{y}^2
     \right], \nonumber
\end{eqnarray}
with the effective parameter $\kappa = J K/D^2$. Note that we have rewritten the expression in such a way that the squares are completed. This representation shows the chirality induced by the N\'eel DMI and the existence of an effective anisotropy.
For large $\kappa > \kappa_{\mathrm{c}} \approx \pi^2/8\approx 1.23$ the model is dominated by the uniaxial anisotropy and the ground state corresponds to ferromagnetic domains oriented along $\pm\hat{\vect{y}}$. In the following, we will focus on the region $\kappa < \kappa_{\mathrm{c}}$ where the ground state is a helical configuration. 

\begin{figure}[tb]
    \centering
                    {\includegraphics[width=\textwidth]{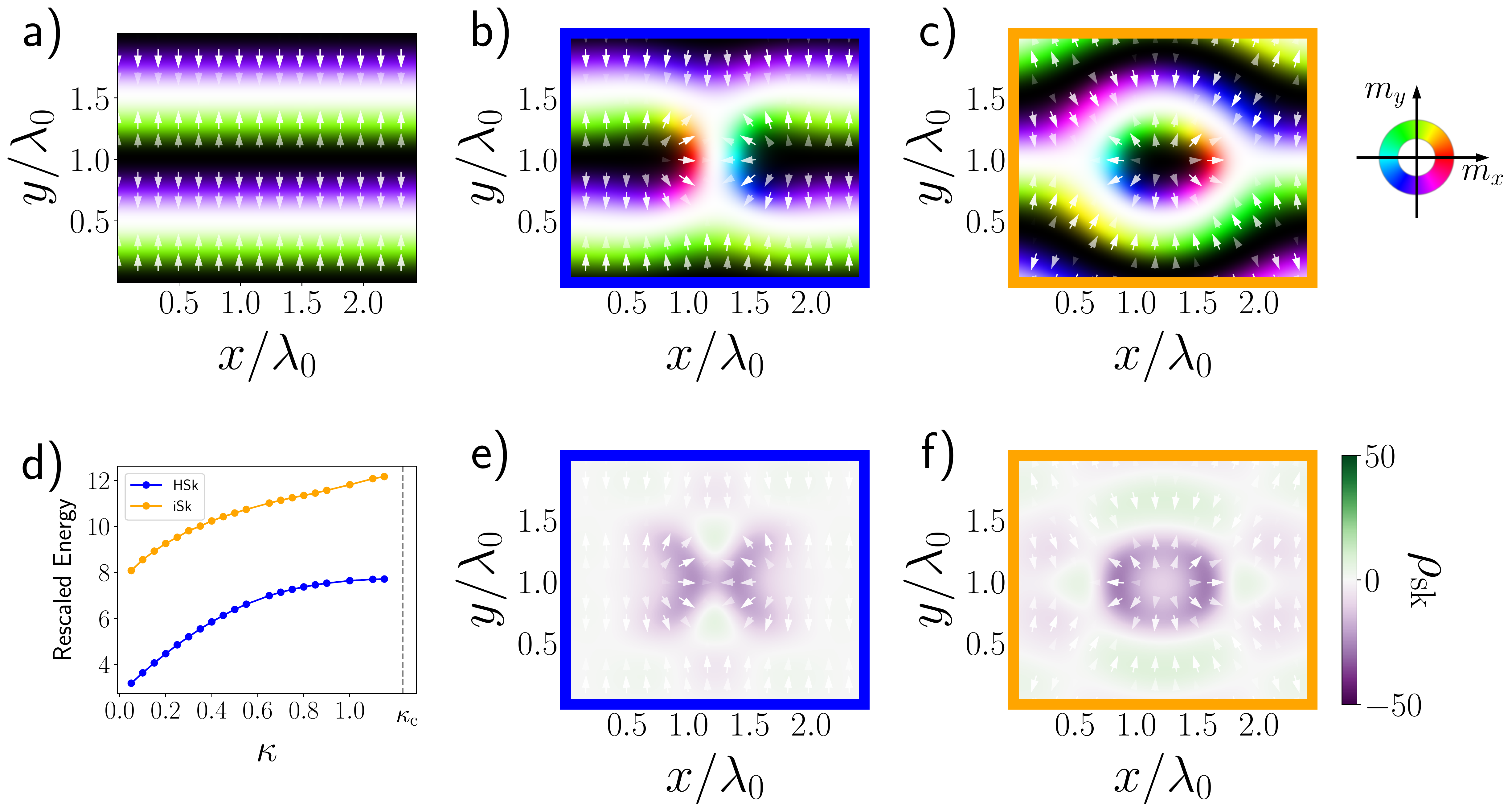}}
\caption{(Meta-) stable magnetic states in helimagnets. a)-c) show the magnetic configurations of a) the helical ground state, b) for the H-shaped skyrmion (HSk) and c) for the interstitial skyrmion (iSk) for $\kappa = 0$. The out of plane component is shown by the contrast $m_z=1$ ($m_z=-1$) corresponds to white (black). The different colours encode the in-plane component of the magnetization as shown by the colour wheel on the right and indicated additionally by the small white arrows.
    d) Rescaled energy as a function of the effective anisotropy $\kappa$ for the two different metastable localized topological states and their skyrmion number density $\rho_{\mathrm{sk}}$ distribution (encoded in colour) is shown in e) and f), respectively. The white arrows indicate the in-plane component of the magnetization at this point, to better compare with the corresponding panels above.
}
    \label{fig:SkyrmionType}
\end{figure}

\subsection{Helical ground state}
\label{subsec:Groundstate}

For $\kappa < \kappa_{\mathrm{c}}$ the ground state of the model of Eq.~(\ref{eq:EnergyRep}) is given by a helical state of the form
\begin{equation}
\label{eq:Helix}
\vect m (\vect r) = \cos\theta(\vect{r}- \vect{r}_{0})\, \hat \vect{q} + \sin\theta(\vect{r}- \vect{r}_{0})\, \hat{\vect{z}}.
\end{equation}
Here the normalized vector $\hat{\vect q}$ characterizes the orientation of the helix, the function $\theta(\vect r)$ describes the profile of the helix, and 
$\vect{r}_{0}$ its spatial displacement. As the considered model, Eq.~(\ref{eq:EnergyRep}) is translational invariant, in the following we will choose $\vect r_0=0$.
In the absence of anisotropy ($\kappa = 0$), the minimal energy configuration is described by the profile function 
\begin{equation}
\label{eq:0Sol} 
\theta (\vect r) \equiv \theta_{\mathrm{h}}^0(\vect{r}) = \hat {\vect{q}} \cdot \vect{r}.
\end{equation}
This corresponds to a helix with wavelength $\lambda_0 = 2\pi $ in the rescaled units. Notice that no orientation of the helix is favoured, i.e.\ $\hat{\vect q}$ is only specified to be in the $xy$ plane.
In the presence of a small easy-axis anisotropy ($0 < \kappa < \kappa_{\mathrm{c}}$), the helical wave vector is favoured to be parallel to the easy axis, i.e.\ $\hat {\vect q} \, ||\, \hat{\vect{y}}$. In this case, the lanes are oriented along the $\hat{\vect{x}}$-direction, see Fig.~\ref{fig:SkyrmionType}a) and the profile function only depends on the $y$-coordinate.~\footnote{Note that a perpendicular magnetic anisotropy favours no particular orientation of the helix, and thus potentially results in current-driven reorientations of the helix direction at boundaries and impurities.~\cite{Everschor2012a, Hals2018a, Masell2020}}
Furthermore, the helix deforms and its profile is given by
\begin{equation}
\label{eq:CSol} 
\theta(\vect{r}) \equiv \theta_{\mathrm{h}}(y)= \textrm{am}\left[\sqrt{2\kappa c} \, y,\frac{1}{c}\right],
\end{equation}
where $\textrm{am}(u,m)$ is the Jacobi amplitude.~\cite{Ezawa2011,Abramowitz1972} The constant $c$ characterizes the conserved linear momentum corresponding to the translational symmetry along $\hat{\vect{y}}$-direction.\footnote{The conserved linear momentum $c = (\partial_{y}\theta)^2/(2\kappa) + \sin^2\theta$ is independent of the coordinate $y$.} It is bound by $1< c< 1 + 1/(2\kappa)$ due to the DMI-induced open boundary condition $\partial_y \theta_{\mathrm{h}}(y) = 1$ while enforcing that at somewhere along the helix the magnetization is in-plane.\footnote{The non-linearity of the solution Eq.~\ref{eq:CSol} is an obstacle to obtain an approximate analytical expression for the $c$ dependence on $\kappa$ that minimizes the energy density.} The natural wavelength of the distorted helix is $\lambda_{\mathrm{h}} = 4 \, \textrm{F}(\pi/2,1/c)/\sqrt{2\kappa c}$, where $\textrm{F}(\pi/2,1/c)$ is the elliptic integral of the first kind.~\cite{Ezawa2011,Abramowitz1972}
As a disclaimer, note that contrary to the ferromagnetic state, the helical state is very much affected by geometrical and boundary effects as well as impurities. Thus, the precise helical structure and its wavelength is, in general, quite complicated for confined samples.

Analytically, we can consider two limits for infinite samples i) $\kappa =0$ and ii) $\kappa \rightarrow \kappa_{\mathrm{c}}$.
The limit $\kappa \rightarrow 0$ corresponds to $c \rightarrow \infty$ for which $\textrm{F}(\pi/2,1/c) \rightarrow \pi/2$ and $\sqrt{2 \kappa c} \rightarrow 1$. With this, one obtains back the helical configuration with the profile described in Eq.~(\ref{eq:0Sol}). 
In the limit $\kappa \rightarrow \kappa_{\mathrm{c}}$ the constant $c$ approaches 1, the wavelength $\lambda_{\mathrm{h}}$ diverges and the magnetic configuration described by Eq.~(\ref{eq:Helix}) turns into a single domain wall with 
$\theta_{\mathrm{h}}(y) = \arccos\left(\tanh\sqrt{2\kappa}y\right)$.

\begin{figure}[tb]
    \centering
     {\includegraphics[width=1.\textwidth]{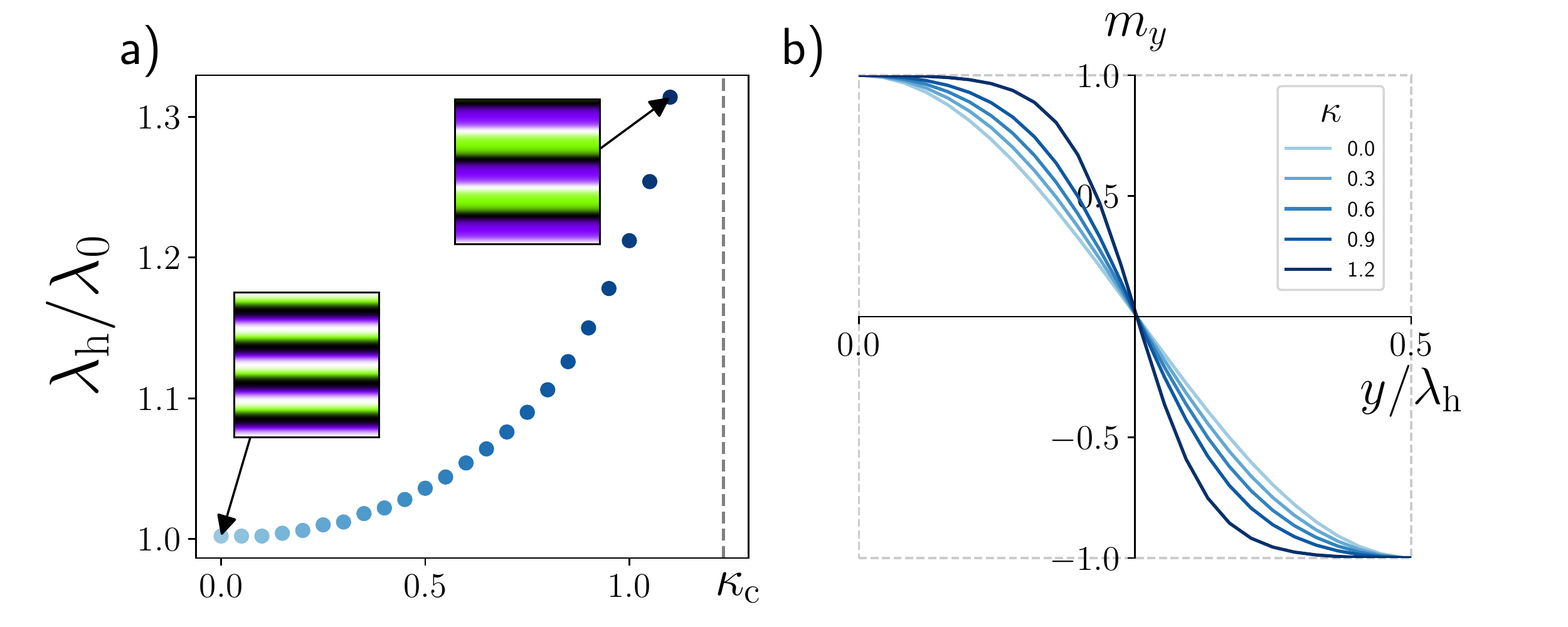}}
\caption{
Characteristics of the helical state as a function of the effective parameter $\kappa$; 
a) The helical wavelength increases as a function of $\kappa$ . Insets show a section of the helical state for $\kappa=0$ and close to the critical value $\kappa_{\mathrm{c}}$ indicated by the grey dashed line.  
Besides the wavelength also the profile of the helices changes as a function of $\kappa$ which is shown explicitly in panel b). The magnetization component along the helical wave vector, here $m_y$ changes from a regular cosine behavior for $\kappa=0$ towards a domain wall-like profile for $\kappa \rightarrow \kappa_{\mathrm{c}}$.}
    \label{fig:helicalprofiles}
\end{figure}

Our numerical results obtained by simulations with \texttt{MuMax}\textsuperscript{3} are shown in Fig.~\ref{fig:helicalprofiles}, for details see App.~\ref{app:Mumax}.
 In Fig.~\ref{fig:helicalprofiles}a) we show the wavelength dependence on the effective parameter $\kappa$ and in Fig.~\ref{fig:helicalprofiles}b) the profiles of the helices for different values of $\kappa$. The numerical simulations reproduce the main qualitative features from the analytics: i) for $\kappa \rightarrow 0$, the domain wall profile obtains a regular cosine pattern as described by Eq.~\ref{eq:0Sol} and ii) that for $\kappa\rightarrow \kappa_{\mathrm{c}}$, the wavelength diverges and the profile converges to that of a domain wall.

\subsection{Skyrmions in the helical ground state}
It has been shown that on top of the helical ground state, localized metastable topological magnetic structures exist~\cite{Ezawa2011, Muller2017}. In a ferromagnetic state,
a local configuration with unitary topological winding number
\begin{equation}
    N_{\mathrm{sk}} =  \frac{1}{4\pi} \int_{\Omega} \rmd^2 r \, \rho_{\mathrm{sk}}\equiv
    \frac{1}{4\pi} \int_{\Omega} \rmd^2 r \, \vect m \cdot \left( \partial_x \vect m \times \partial_y \vect m \right),
\end{equation}
 is typically associated to a magnetic skyrmion.
 In the helical background there are two distinct configurations associated with the topological charge $N_{\mathrm{sk}} = 1$: \emph{interstitial skyrmions} (iSk) and \emph{\enquote{H-shaped} skyrmions} (HSk), see Fig.~\ref{fig:SkyrmionType}b) and c). The former resemble skyrmions in ferromagnets, around which the helical background bends.~\cite{Muller2017} The latter are bound meron pairs, where each meron has a half-integer topological charge.~\cite{Ezawa2011,Muller2017,Gao2019}. Their corresponding skyrmion densities $\rho_{\mathrm{sk}}$ are shown in Fig.~\ref{fig:SkyrmionType}e) and f). 
In agreement with Ref.~\cite{Muller2017}, we find that HSk are energetically favourable compared to iSk, see Fig.~\ref{fig:SkyrmionType}d).

\section{Current-driven dynamics in helical states}
\label{sec:dynamics}
In the following, we discuss the current-induced dynamics of helical states subject to spin-transfer torques,~\cite{Gilbert2004,Li2004} which is well described in the framework of the Landau-Lifshitz-Gilbert (LLG) equation
\begin{equation}\label{eq:LLGST}
    \dot{\vect m}= - \vect m \times \vect B_{\mathrm{eff}} -\left(\vect{v}_{\mathrm{s}} \cdot \nabla\right) \vect m +\alpha \vect{m} \times \dot{\vect{m}} 
    +\beta \vect{m} \times ( \vect v_{\mathrm{s}} \cdot \nabla) \vect m.
\end{equation}
Here, $\vect B_\mathrm{eff} = -\delta E[\vect m] / \delta \vect m$ and $\alpha, \beta$ are the damping parameters. Time has been rescaled by $\tau= J M_{\mathrm{s}}/\gamma D^2$ where $\gamma$ is the (positive) gyromagnetic ratio and $M_{\mathrm{s}}$ is the saturation magnetization. The strength of the spin-transfer torque is characterized by the velocity $\vect{v}_{\mathrm{s}}$ which is proportional to the electrical current density $\vect{j}_{\mathrm{e}}$ as
$\vect{v}_{\mathrm{s}} = - \vect{j}_{\mathrm{e}} P \mu_{\mathrm{B}}/(e \gamma D(1+\beta^2))$,
where $P$ is the polarisation, $\mu_{\mathrm{B}}$ is the Bohr magneton, and $e > 0$ is the electron charge. 

To analyze the effects of spin-transfer torques, which couple only to gradients of the magnetic structure, one has to consider several regimes. In a real sample, pinning forces favour static equilibrium configurations while spin-torques inject energy into the system and drive the system away from equilibrium. Their interplay divides the dynamical phase-space into three regimes ordered by increasing driving strength in 
i) $v_{\mathrm{s}} < v_{\mathrm{pin}}$ the magnetic configuration is static; 
ii) $v_{\mathrm{pin}} < v_{\mathrm{s}} < v_{\mathrm{C}}$ the magnetic configuration evolves controllably; iii) $v_{\mathrm{s}}>v_{\mathrm{C}}$ the magnetic configuration is unstable and there is no long-range magnetic order. 
While the strengths of $v_{\mathrm{pin}}$ and $v_{\mathrm{C}}$ depend on the particular system, more specifically on the magnetization configuration, geometry and potential disorder, a few general remarks can be made.
In homogeneous systems an example of dynamics in region (ii) is the steady translation of rigid configurations, which can be described by a  Thiele equation~\cite{Thiele1973,Everschor2011}. For non-homogeneous systems the magnetization dynamics becomes more complex ~\cite{Everschor-Sitte2016,Tatara2004,Sitte2016,Muller2016a,Masell2020a,Rodrigues2020} and gives rise to interesting phenomena such as the shedding of topological textures.~\cite{Sitte2016,Muller2016a,Everschor-Sitte2016,Rodrigues2020}

\subsection{Instabilities of the helical state}
\label{subsec:CurrentPert}
We consider first the \emph{pure} helical state with open boundary conditions in the absence of disorder and at zero temperature. In this case, the magnetization gradient only couples to the $\hat{\vect{y}}$-component of the driving. Furthermore, 
$v_{\mathrm{pin}}=0$, and spin-transfer torques lead to a global translation of the helices along the $\hat{\vect{y}}$-direction. For drivings above a strength $v_{\mathrm{C}}\equiv v_{\mathrm{W}}$ the helical state becomes unstable, and the magnetisation may start to precess, leading to a Walker-breakdown instability~\cite{Masell2020}, similar to the one for a domain wall.\footnote{In the absence of anisotropy $v_{\mathrm{W}}=1$ and the Walker breakdown of the helix is the transition from driven conical to polarized.}

In a typical experiment, however, the helical phase is very sensitive to geometry, boundary effects and disorder. Therefore, $v_{\mathrm{pin}}$ can only be determined for the specific considered configuration. Below we will consider a regime where we prevent the global translation of the helices along the $\hat{\vect{y}}$-direction by pinning the helices on the boundaries, as the key idea is to work in a regime where only local structures, i.e. the skyrmions, move along the tracks and not the full helical system, see Sec.~\ref{sec:motion}. 
For such pinned helical states with wave vector along $\hat{\vect{y}}$, the $y$-component of the drive will deform the profile of the helices while the 
$x$-component couples with any perturbations of the helical background. 
It turns out that this pinned helix regime splits further into two cases, one in which the applied drive is small in the sense that perturbations only act locally on a size much smaller than the helical wavelength $\lambda_{\mathrm{h}}$ and one where perturbations are on the order of $\lambda_{\mathrm{h}}$. The latter allows to change the structure of the helical state and produce metastable local excitations such as the HSk and iSk. We denote the drive above which a local excitation modifies the helical state by $v_{\mathrm{c}}$.

Analytically we can obtain an estimation of $v_{\mathrm{c}}$ for $\kappa \lesssim \kappa_{\mathrm{c}}$, where the helical state is distorted in such a way that it can be interpreted as a sequence of ferromagnetic domains periodically separated by domain walls, see inset of Fig.~\ref{fig:helicalprofiles}a) for $\kappa$ close to $\kappa_{\mathrm{c}}$.
The lowest excitation modes of such a state can be interpreted as a local translation of a domain wall with a rigid profile~\cite{Rodrigues2018} within the length scale of a helical period to not shift the full helical state.
Therefore, we consider the magnetization locally given in terms of the spherical angles as 
\numparts
\begin{eqnarray}
\vect m (\vect r) = \sin \theta(\vect r) \hat{\vect{z}} + \cos\theta(\vect r) \left[\cos\phi(\vect r) \, \hat{\vect{x}} + \sin\phi(\vect r)\, \hat{\vect{y}}\right], \\
\fl \text{with}\ \ \theta(\vect{r}) = \arccos\left(\tanh\left(\sqrt{2\kappa}(y - y_{0}(x))\right)\right),\ \mathrm{and} \  \ \phi(\vect{r}) = \frac{\pi}{2} + \varphi(x),
\end{eqnarray}
\endnumparts
in a region $y - y_{0}\ll \lambda_{\mathrm{h}}/2$. Here $y_0 (x)$ and $\varphi (x)$ describe the fluctuations along the tracks that might arise due to some local perturbation in this region.
We consider the magnetization to be static everywhere else corresponding to the helical background. In this approximation, we claim that the energy associated to perturbations of the helical background is equivalent to the energy associated to a perturbation of a domain wall in a confining potential, i.e.\ $E_{\mathrm{cp}}[\vect m(\vect{r})] \approx E[\vect m_{DW}(\vect{r})] +\int dx\, 2y_{0}^2/\lambda_{\mathrm{h}}^2$, and yields
\begin{equation}
\label{eq:approxE}
\fl E[\vect m(\vect{r})] \approx \int dx \left( \frac{1}{\sqrt{2\kappa}}(\partial_{x}\varphi)^2 + \sqrt{2\kappa}(\partial_{x}y_{0})^2 - \frac{\pi}{2}\varphi\partial_{x}y_{0} + \left(\frac{\pi}{2} + \sqrt{2\kappa}\right)\varphi^2 + \frac{2y_{0}^2}{\lambda_{\mathrm{h}}^2}\right).
\end{equation}
The last two terms correspond to confining potentials that fix the helicity and position of the domain wall, respectively. Notice that the potential for 
$\varphi$ is positive for $\kappa < \kappa_{\mathrm{c}}$, and the constraining potential for $y_{0}$ vanish as $\kappa$ approaches the critical value.

For a current applied along the $\hat{\vect x}$-direction and in the absence of non-adiabatic damping ($\beta=0$), from Eq.~(\ref{eq:LLGST}) and Eq.~(\ref{eq:approxE}) we deduce the linearized equations of motions for $y_0$ and $\varphi$,
\numparts
\begin{eqnarray}
\partial_{t}\varphi = \left(v_{x} + \frac{\pi}{2}\right)\partial_{x}\varphi - \sqrt{2\kappa} \partial_{x}^2 y_{0} + \frac{2}{\lambda_{\mathrm{h}}^2} y_{0} - \sqrt{2\kappa} \alpha \partial_{t}y_{0},\\
\partial_{t}y_{0} = \left(v_{x} + \frac{\pi}{2}\right)\partial_{x}y_{0} + \frac{1}{\sqrt{2\kappa}}\partial_{x}^2\varphi - \left(\frac{\pi}{2} + \sqrt{2\kappa}\right)\varphi + \frac{1}{\sqrt{2\kappa}}\alpha\partial_{t}\varphi.
\end{eqnarray}
\endnumparts
Analysing the stability of the system, in the limit of small damping $\alpha$, 
we find that for $\kappa \lesssim \kappa_{\mathrm{c}}$ perturbations in the system damp out for drives below
\begin{equation}\label{eq:criticdrive}
v_{c}^{\kappa \lesssim \kappa_{\mathrm{c}}}\approx \sqrt{\frac{2}{\lambda_{\mathrm{h}}^2} + \frac{\pi}{2} + \sqrt{2\kappa}} - \frac{\pi}{2}.
\end{equation}
which in the limit $\kappa=\kappa_{\mathrm{c}}$ yields $v_{\mathrm{c}}^{\kappa_{\mathrm{c}}}\approx 0.20$. Moreover, in this approximation, one notices that as $\kappa \rightarrow 0$ the helical state becomes unstable at any non-zero drive, $|v_{s}| >0$.~\cite{Masell2020}
Above this critical current, any perturbation tends to grow exponentially. Depending on the detailed energy landscape, such perturbations span a great variety of possible magnetic configurations and may allow for the creation of metastable states. This regime, for example, allows for the creation of HSk or iSk and will be explored in  Sec.~\ref{sec:shedding}. 

\begin{figure}[tb]
    \centering
                {\includegraphics[width=\textwidth]{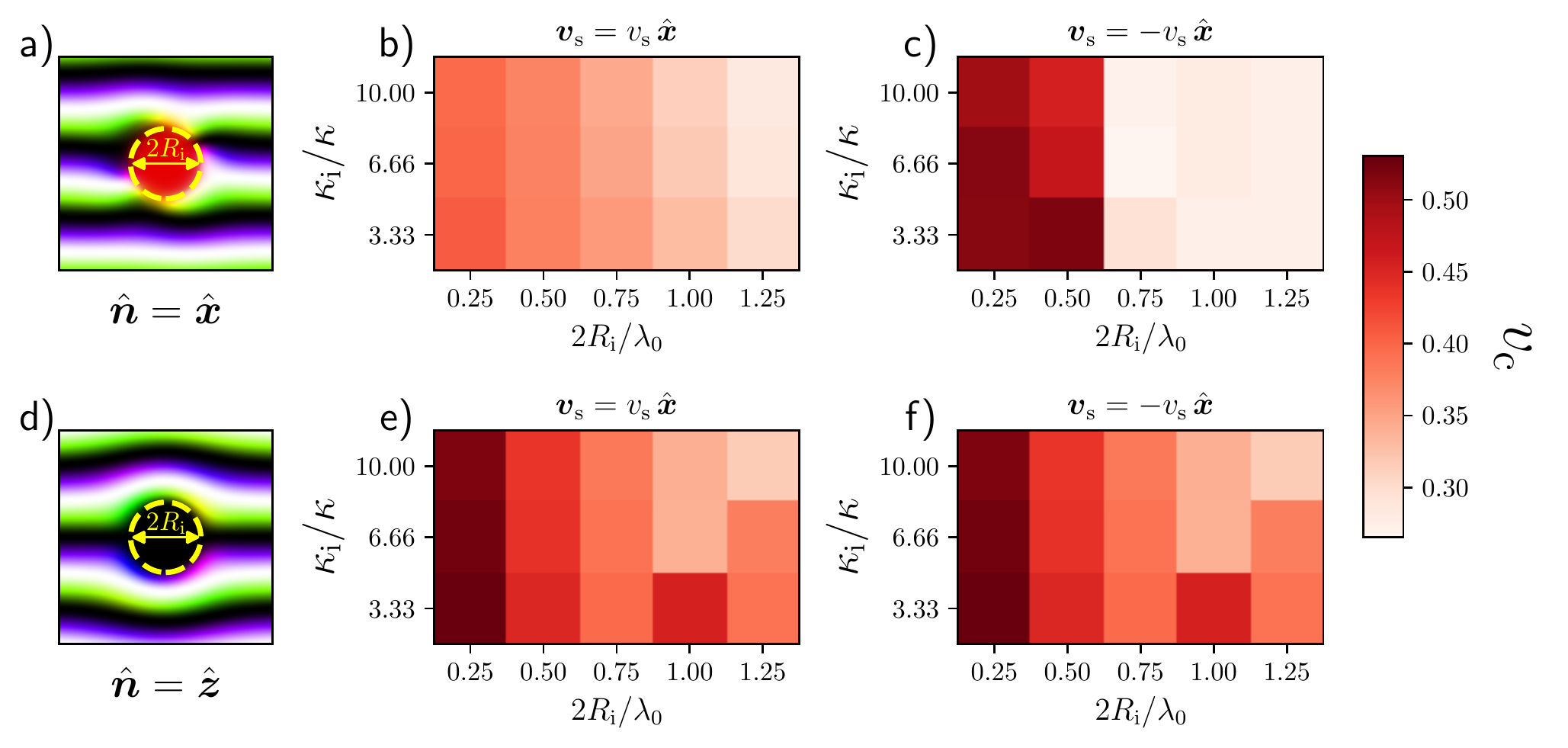}}
\caption{Current-induced instability of the helical phase due to an inhomogeneity. a) and d) show an example magnetisation configuration around the impurity region for $\hat{\vect{n}}= \hat{\vect{x}}$ and $\hat{\vect{n}}= \hat{\vect{z}}$, respectively. 
    b), c), e), f) show the critical drives $v_{\mathrm{c}}$ as a function of inhomogeneity radius and relative anisotropy strength $\kappa_{\mathrm{i}}/\kappa$. 
 For all panels $\kappa\approx0.31$, $\alpha=0.05$ and $\beta=0$ and in a) and d) we used $2 R_{\mathrm{i}} = \lambda_0$ and $\kappa_{\mathrm{i}}/\kappa=10$.  We note that the graphs e) and f) are identical due to the mirror symmetry along $\hat{\vect{x}}$ for the case of an impurity along $\hat{\vect{z}}$.
}
    \label{fig:SizeStrengthPhaseDiagram}
\end{figure}

To numerically study the critical current density above which the helical state significantly deforms, we consider an impurity as a circular region of radius $R_{\mathrm{i}}$ with an easy-axis anisotropy along $\hat{\vect{n}}$ perpendicular to $\hat{\vect{y}}$. The energy contribution for the impurity to the functional in Eq.~(\ref{eq:EnergyRep}) is given by
\begin{equation}
\label{eq:inhomogeneity}
E_{i}[\vect{m}(\vect{r})] = -\int \rmd^2 r \, \kappa_{\mathrm{i}} (\hat{\vect{n}}\cdot\vect{m})^2 \, \Theta\left(R_{\mathrm{i}}^2 - (x-x_{i})^2 - (y-y_{i})^2\right), 
\end{equation}
where $\kappa_{\mathrm{i}}$ is the strength of the local easy-axis anisotropy along $\hat{\vect{n}}$, $\Theta(x)$ is the Heaviside step function being 1 for $x\geq0$ and $0$ for $x<0$, the coordinates $x_{i},y_{i}$ are the position of the inhomogeneity region. The inhomogeneity leads to a local deformation of the helix in this region, see Fig.~\ref{fig:SizeStrengthPhaseDiagram}a) and d).
When applying a current, the deformation enhances and above a certain drive $v_{\mathrm{c}}$ does no longer allow for a statically stable solution. In Fig.~\ref{fig:SizeStrengthPhaseDiagram} we show the results obtained with \texttt{MuMax}\textsuperscript{3} for systematically computing the critical drive $v_{\mathrm{c}}$ as a function of the ratio of the local and global anisotropy strengths $\kappa_{\mathrm{i}}/\kappa$, the radius $R_{\mathrm{i}}$ and for $\hat{\vect{n}}$ along $\pm \hat{\vect{x}}$ and $\pm \hat{\vect{z}}$. The results shown in Fig.~\ref{fig:SizeStrengthPhaseDiagram} are computed for $\kappa=0.31$, $\alpha=0.05$ and $\beta=0$, for details see App.~\ref{app:Mumax}. 

The presence of the inhomogeneity modifies the magnetization locally as the magnetization inside the inhomogeneity tends to rotate towards $\hat{\vect{n}}$ see Fig.~\ref{fig:SizeStrengthPhaseDiagram}a) and, d). The stronger the inhomogeneity the more the magnetization aligns with the anisotropy direction of the impurity, and the domain wall stripes are pushed out of the inhomogeneity region. As the initial perturbation is a bit stronger for a larger $\kappa_{\mathrm{i}}/\kappa$, the critical current decreases as a function of $\kappa_{\mathrm{i}}/\kappa$.
The most important factor for the deformation of the helical state, however, is the size of the inhomogeneity, which essentially determines the curvature of the domain wall.
We find that the critical drive generally decreases with an increasing size since a bigger impurity typically produces a large initial perturbation that couples more strongly to the spin-transfer torque.

For the impurity region with $\hat \vect n = \hat \vect z$, the set-up is still axial symmetric and therefore a current along $+ \hat \vect x$ or $-\hat{\vect{x}}$-direction yields the same critical drive. For $\hat \vect n = \hat \vect x$, the impurity breaks the axial symmetry and it induces the magnetization to align perpendicularly to the plane defined by the helical configuration, generating an extra gradient component. Therefore, reversing the current direction alters the size of the critical current. 

%%%%%%%%%%%%%%%%%%%%%%%%%%%%%%%%%%%%%%%%%%%%%%%%%%%%%%%%%%%%%%%%%%%%%%%%%
\subsection{Current-Driven Motion of Skyrmions in the Helical Background}
\label{sec:motion}
%%%%%%%%%%%%%%%%%%%%%%%%%%%%%%%%%%%%%%%%%%%%%%%%%%%%%%%%%%%%%%%%%%%%%%%%%

A key prerequisite of spintronic-based racetrack memories is the possibility to move the rigid localized magnetic configurations by applying a current. To obtain the motion of the skyrmions in the helical background, we consider that in the current-driven steady motion 
the magnetization is given by $\vect{m}(\vect{r},t) \equiv \vect{m}(\vect{r} - \vect{v}_{\mathrm{s}}t)$. Given this assumption and projecting Eq.~(\ref{eq:LLGST}) along $\vect m\times\partial_{x} \vect m$ and $\vect m \times\partial_{y}\vect m$ leads to the Thiele equation~\cite{Thiele1973,Everschor2011}
\begin{equation}
\label{eq:forces}
\vect{\mathcal{G}}\times(\vect{v} - \vect{v}_{s}) +\mathcal{D} ( \alpha \vect{v} -\beta \vect v_{\mathrm{s}})- \vect{F} = \vect{0},
\end{equation}
where $\vect{\mathcal{G}} = 4\pi \hat{\vect{z}}$ is the gyrovector, $\mathcal{D}_{ij} = \int \rmd^2 x \partial_{i}\vect m \cdot \partial_{j}\vect m$ is the dissipative tensor, and $\vect{F} = \partial_{x} E[\vect m(\vect{r})]\hat{\vect{x}} + \partial_{y} E[\vect m(\vect{r})]\hat{\vect{y}}$ is a force due to space variations of the energy density functional. 
The first term in the equation is responsible for the skyrmion Hall angle, it produces a movement of the skyrmion that is perpendicular to the perturbation. The second term is a dissipative term whose direction of motion depends on the shape of the skyrmion. 
The third term pushes the skyrmion towards a stable position in the energy landscape. The helical background has a significant impact on the last two terms by producing a non vanishing $\vect{F}$ and deforming the skyrmion as it is displaced from a path of minimal energy. The helical state also confines the motion of the skyrmions along the helical tracks, i.e.\ $\vect{v} \parallel \hat{\vect{x}}$ in a steady motion. Solving Eq.~(\ref{eq:forces})
for the velocity of the skyrmion thus yields for the steady motion 
\begin{equation}
\label{eq:skyrmionspeed}
\vect{v} = \left(\frac{4\pi (\vect{v}_{s}\cdot\hat{\vect{y}})}{\alpha \mathcal{D}_{xx}} + \frac{\beta}{\alpha}\vect{v}_{s}\cdot \frac{(\mathcal{D}_{xx} \hat{\vect{x}} + \mathcal{D}_{xy} \hat{\vect{y}})}{\mathcal{D}_{xx}}\right)\hat{\vect{x}}.
\end{equation}
For small drives where this equation holds, the velocity grows linearly with the applied current.
Note that Eq.~(\ref{eq:skyrmionspeed}) holds for all metastable magnetic structures in the helical background, including the HSk and iSk. Furthermore, in the absence of non-adiabatic spin-transfer torques, i.e.\ $\beta=0$, only a current along $\hat \vect y$ will induce a motion of the skyrmion.

\begin{figure}[tb]
    \centering
    \includegraphics[width=\linewidth]{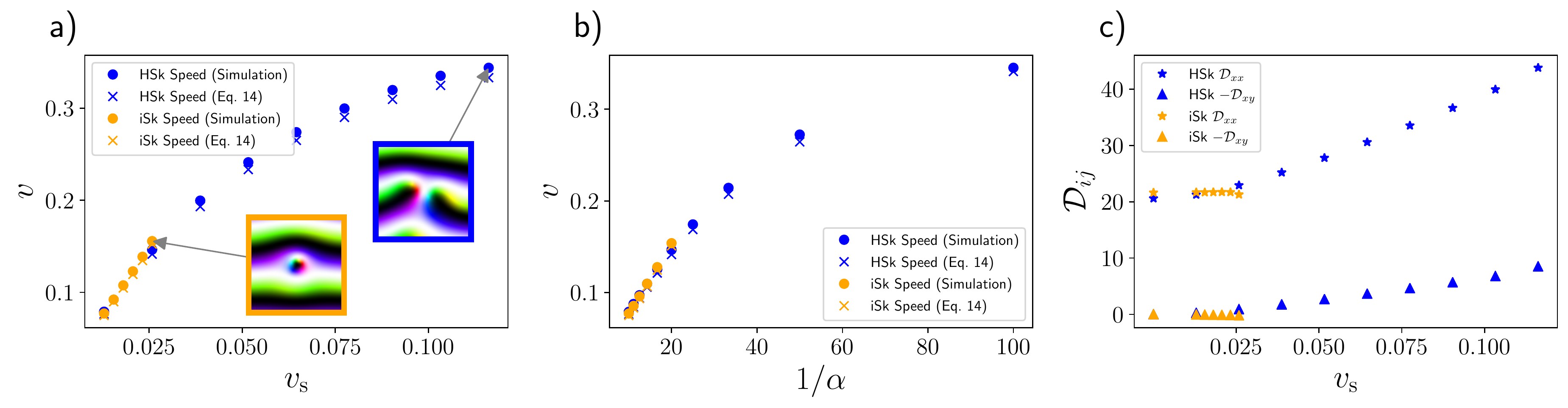}
    \caption{Properties of HSk and iSk subject to an applied drive $v_{\mathrm{s}}$ along the $\hat{\vect{y}}$-direction. a) and b) show the skyrmion speed $v$ as a function of  driving strength $v_{\mathrm{s}}$ and inverse damping $1/\alpha$, respectively. For small drives the corresponding relation is linear and becomes sub linear for larger drives. High skyrmion speeds can be obtained in the regime of small damping and large drives. Note that because of the HSk being more stable it generally allows for higher drives than the iSk. For large drives both skyrmions deforms as exemplarily shown in the insets.
 The dotted data points have been obtained by numerical simulations and the cross-shaped data points have been obtained using Eq.~\ref{eq:skyrmionspeed} while micromagnetically computing the dissipative tensor $\mathcal{D}$, the drive dependence of which is shown in c). 
 For this plot we used $\kappa \approx 0.31$, and $\beta=0$. In a) and c) $\alpha=0.1$, and in b) $v_{\mathrm{s}} \approx \num{2.1e-3}$.
        }%
    \label{fig:CurrentSpeed}
\end{figure}

For larger drives, where the skyrmion tends to move away from the local minimum given by the helical track, the skyrmion gets deformed with a deformation that depends on the strength of the applied current. By means of micromagnetic simulations we show that, in this case, the relation of the skyrmion speed and the drive becomes sublinear, see Fig.~\ref{fig:CurrentSpeed}a). 
In Fig.~\ref{fig:CurrentSpeed} we show the speed and deformation of the HSk and iSk skyrmion as a function of the applied drive and inverse damping in the absence of non-adiabatic spin-transfer torques obtained by micromagnetic simulations as well as the results of Eq.~(\ref{eq:skyrmionspeed}) when computing the dissipative tensor  $\mathcal{D}$ micromagnetically, shown in panel c). High-speed skyrmion motion arises for high drives and low damping. We also find that, as expected, the HSk is more robust, and thus allows for higher velocities compared to the iSk which annihilates when pushed too much away from its minimal energy position. 

%%%%%%%%%%%%%%%%%%%%%%%%%%%%%%%%%%%%%%%%%%%%%%%%%%%%
\section{Shedding of Skyrmions in the Helical State}
\label{sec:shedding}
%%%%%%%%%%%%%%%%%%%%%%%%%%%%%%%%%%%%%%%%%%%%%%%%%%%%

\begin{figure}[tb]
    \includegraphics[width=\linewidth]{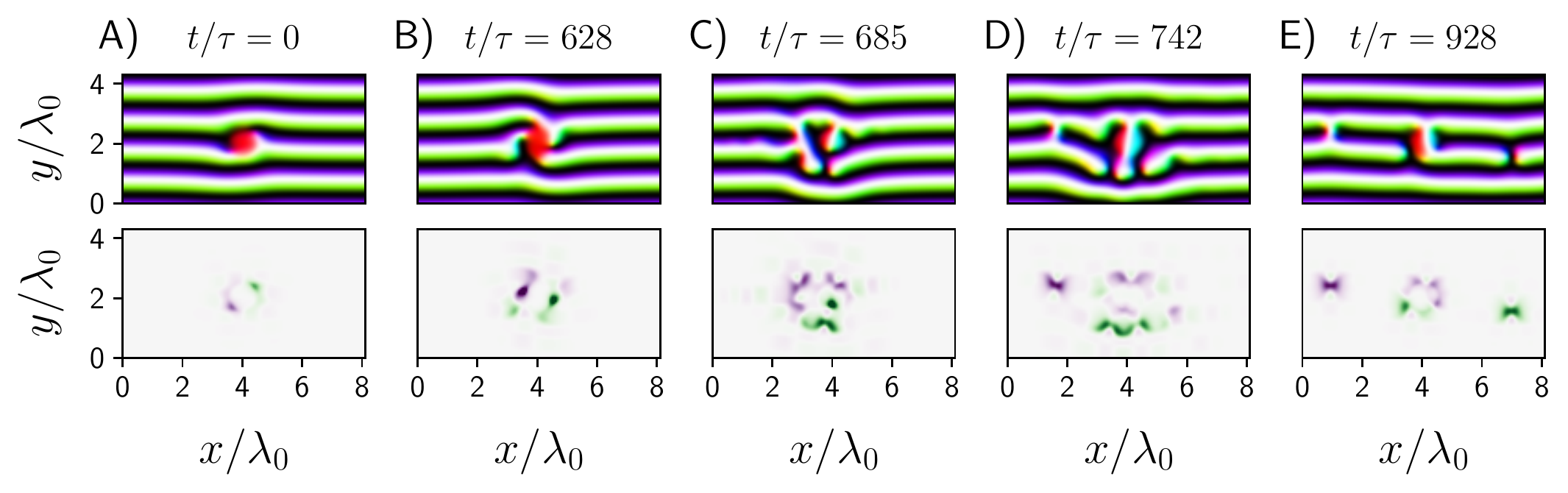}
    \caption{Creation of a Sk/anti-Sk pair in a helical background through the interplay of an inhomogeneity and an applied current. The top (bottom) row shows the magnetization configuration (skyrmion number density) around the impurity at different times during the creation process. Panel e) shows the created HSK/anti-HSk pair. The colour coding is identical to the one used in Fig.~\ref{fig:SkyrmionType}. The material parameters are discussed in App.~\ref{app:Mumax}. 
  }
    \label{fig:singleshedding}
\end{figure}

\begin{figure}[tb]
    \includegraphics[width=\linewidth]{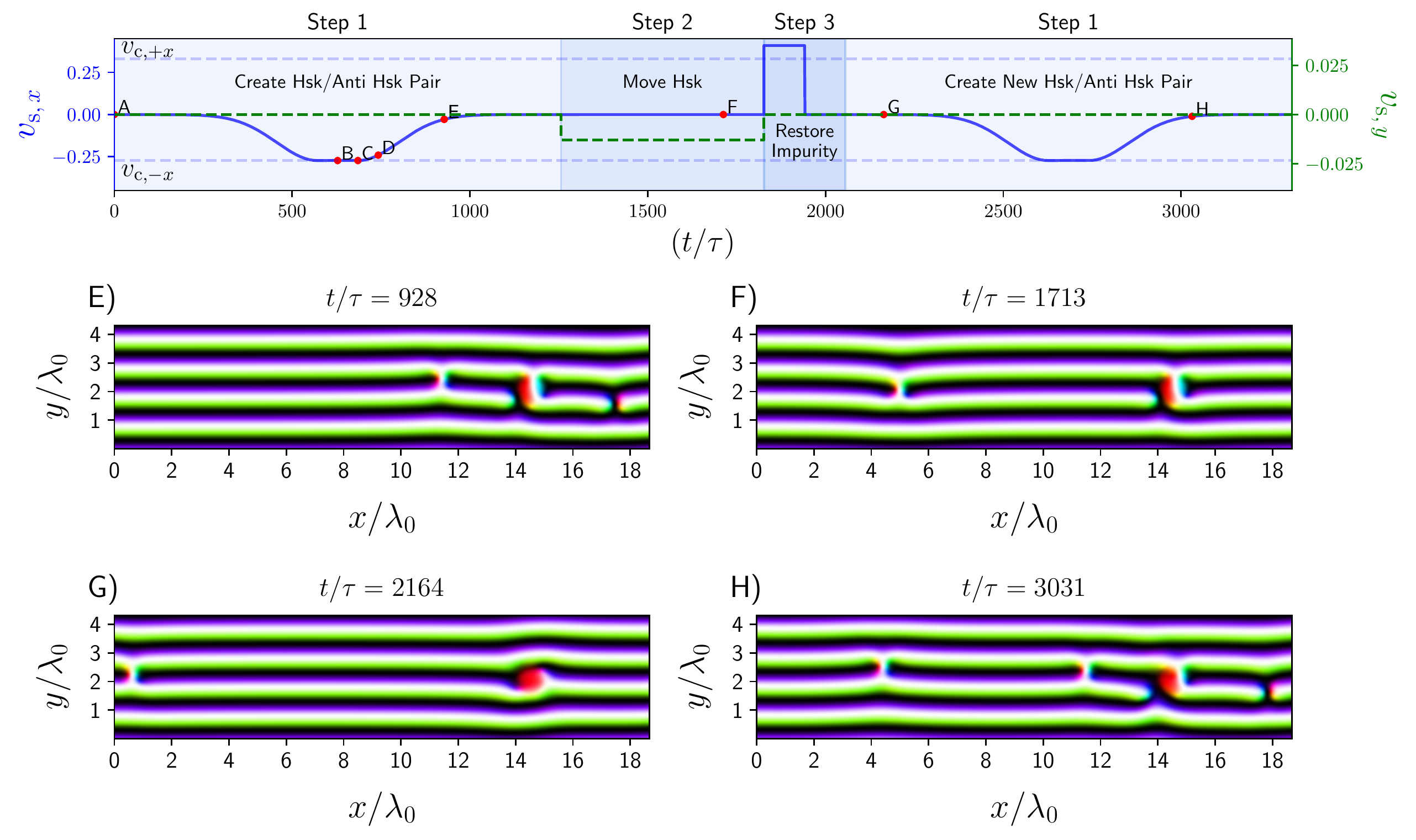}
    \caption{All electrical three-step HSk creation protocol. a) shows the applied current pulses as a function of time. The labels A-H mark the times where we show the corresponding magnetization configurations. The details of the HSk/anti-HSk creation (A-E) are shown in Fig.~\ref{fig:singleshedding}. In b) we first repeat the configuration of the HSk/anti-HSk after its creation (E) and then we focus on the HSk that has been moved away from the impurity region with a pulse in $\hat{\vect{y}}$-direction. Under the influence of such a pulse the anti-HSk moves in the opposite direction, i.e.\ out of the focus region here.   
    In c) we show the result of step 3, the restored impurity (G), which then allows for the production of another HSk/anti-HSk pair. 
    The material parameters are discussed in App.~\ref{app:Mumax}.  }
    \label{fig:PulsedShedding}
\end{figure}

For the controllable injection of skyrmions we consider the design with a localized inhomogeneity in the sample, as described by Eq.~(\ref{eq:inhomogeneity}) with $\hat \vect n = \hat \vect x$, $\kappa_{\mathrm{i}} = 10 \, \kappa$ and $2 R_{\mathrm{i}}/\lambda_0 = 1.0$. 
Applying a drive along the $\hat{\vect{x}}$-direction above $v_{\mathrm{c}}$ deforms the magnetization around the impurity, and, 
due to the conservation of topological charge $N_{\mathrm{sk}}$, not only creates a HSk but a HSk/anti-HSk pair on opposite sides of the impurity, see Fig.~\ref{fig:singleshedding}. After the creation of this pair, unfortunately the magnetization configuration around the impurity is in a metastable state which does not immediately allow for another pair creation.
Combining this fact with the current-driven skyrmion motion discussed in Sec.~\ref{sec:motion}, we developed the following all-electrical three-step protocol for creating skyrmions in the helical background, shown in Fig.~\ref{fig:PulsedShedding}:
\begin{itemize}
\item {\bf Step1:} Create a HSk/anti-HSk pair: Starting from the helical state with an impurity, use a current pulse along $-\hat{\vect{x}}$ with strength above the critical current.
\item {\bf Step2:} Move HSk and anti-HSk away from the impurity region: Apply a current along $\hat{\vect{y}}$ to move the skyrmions. Notice that the HSk and anti-HSk move in opposite directions.
\item {\bf Step3:} Restore the impurity configuration: Apply a short pulse of current along $\hat{\vect{x}}$ with a strength above the critical current in that direction.
\end{itemize}
Once the magnetization at the impurity is reinitiated to its lowest energy configuration, the protocol can be repeated to create another Sk/anti-Sk pair. 
We note that the size and duration of the pulses must be tailored for the size and strength of the impurity as discussed in Fig.~\ref{fig:SizeStrengthPhaseDiagram} 
This enhances the control over the skyrmion creation since different impurities will not be able to generate and be reset by the same applied current pulses allowing for precise creation of skyrmions at the desired locations.
Furthermore, the skyrmions are moved with much smaller currents applied along the $\hat{\vect{y}}$-direction, allowing the process of creation and skyrmion transport to be carried out rather independently from each other.

%%%%%%%%%%%%%%%%%%%%%%%%%%%%%%%%%%%
\section{Discussion and Conclusion}
\label{sec:disscon}
%%%%%%%%%%%%%%%%%%%%%%%%%%%%%%%%%%%
We studied the stability and current-driven dynamics of helical states with in-plane uniaxial anisotropy in the presence of perturbations and inhomogeneities. Compared to the ferromagnetic state, the helical state has more degrees of freedom making its physics richer. 
Besides naturally providing tracks for skyrmion motion, advantages of helical phases are that they exist at room temperature in absence of external magnetic fields, and that they are very robust to external perturbations.~\cite{Adams2012,Zhang2017}

As a central result, we developed a theory to explain for which current ranges it is possible to produce and manipulate metastable topological structures in the helical background. Based on these findings we developed an all-electrical three-step protocol that independently allows for the controlled creation and the controllable motion of HSks.
Our results obtained by micromagnetic simulations are supported by analytical calculations where the latter are mainly performed in the limits of small or high in-plane anisotropy. 
For simplicity, we neglected stray fields in our calculations. Generally, we expect their influence not to be too important, as the helical state itself has zero net magnetization. We expect that their main effect is to distort the helical profile mildly, and enhance the stability of the HSk and iSk, as magneto-static interactions favour the formation of twisted structures. Furthermore, our numerical results were obtained in the absence of damping-like spin-transfer torques, i.e.\ $\beta = 0$. For $\beta \neq 0$  we expect the current to couples more effectively to the magnetization and lead to lower critical currents $v_{\mathrm{c}}$ and higher skyrmion speeds, similar as in Ref.~\cite{Rodrigues2020}. Also, all our simulations are performed at zero temperature. Temperature induces perturbations in the well-ordered states, and thus might alter the stability of the twisted magnetic structures as well as lower the critical current densities.

As a final remark, in the absence of a magnetic field, the helical background does not favour HSk over anti-HSks. During the three-step protocol, the anti-HSk was not in our focus, as the idea was to show a protocol to create multiple HSks for a racetrack-like device. But since HSk/anti-HSk pairs are created on the opposite side of the impurity and moved away from the impurity in opposite directions, this protocol allows for information to be created and moved along two directions simultaneously, going even beyond the concept of a HSk-based helical magnetization racetrack memory.

%%%%%%%%%%%%%%%%%%%%%%%%%
\section{Acknowlegdments}
%%%%%%%%%%%%%%%%%%%%%%%%%

We thank O.~Gomonay for fruitful discussions and J.~Sinova for the constant great support.
We acknowledge funding from the German Research Foundation (DFG) under Project Nos.~403233384 (SPP Skyrmionics), 320163632 (Emmy Noether) and 268565370 (TRR 173, project B12). R.~K.~is supported by a scholarship from the Studienstiftung des deutschen Volkes.
J.~M.~is supported by JSPS (project No.19F19815) and the Alexander von Humboldt foundation.

%%%%%%%%%%%%%%%%%%%%%%%%%%%%%%%%%%%
\section{Supplementary Information}
%%%%%%%%%%%%%%%%%%%%%%%%%%%%%%%%%%%
%\subsection{Details on Numerical Simulations}
\label{app:Mumax}

The micromagnetic simulations were performed with \texttt{MuMax}\textsuperscript{3}~\cite{Vansteenkiste2014}. For all simulations we used $M_{\mathrm{s}} = \SI[parse-numbers=false]{10^6}{\ampere\per\metre}$, spin polarization $P=1$, $J = \SI{20}{\pico\joule\per\metre}$ and $D\approx \SI{2.55e-3}{\joule\per\square\metre}$, yielding a pure helical wave length ($K = \SI[parse-numbers=false]{0}{\joule\per\cubic\metre}$) of $\lambda_0 = 2\pi J / D \approx \SI{49.3}{\nano\metre}$ and a characteristic time scale of $\tau= J M_{s}/\gamma D^2 \approx \SI[parse-numbers=false]{0.0175}{\nano\second}$. The choice of the DMI strength is motivated by choosing $D = 2 D_{\mathrm{c}}$, where $D_{\mathrm{c}} = 4 \sqrt{JK/2} / \pi$,~\cite{Rohart2013} i.e. a regime where we found the skyrmions to be metastable. To obtain $D\approx \SI{2.55e-3}{\joule\per\square\metre}$ we used $K = \SI[parse-numbers=false]{10^5}{\joule\per\cubic\metre}$ for which we obtain an effective parameter of $\kappa=JK/D^2 \approx 0.31$.
The effective parameter $\kappa$ is only ever varied by varying $K$, and unless otherwise specified, $K = \SI[parse-numbers=false]{10^5}
{\joule\per\cubic\metre}$.
Furthermore, we chose the damping constants $\alpha= 10^{-3}$ and $\beta = 0$ unless otherwise specified. 
Note that we made sure to adjust the system size to allow for the natural helical solution avoiding geometrical influence of finite size samples.

To pin the helical state in the simulations with applied current, the spins at the top and bottom of the system were frozen.
To compensate for current-induced effects at the sample boundaries, we suppressed the spin dynamics at the boundary by setting a high damping constant at the left and right ends of the system, $\alpha = 5$. 

The proportionality factor between the drive velocity $\vect v_{\mathrm{s}}$ and the electrical current density is $P \mu_{B}/(e \gamma D(1+\beta^2)) 
\approx \SI[parse-numbers=false]{1.29\cdot 10^{-13}}{\square\metre\per\ampere}$.
 A typical used drive value of $v_{\mathrm{s}}=0.025$ to move the skyrmions corresponds to 
 $j_e\approx \SI[parse-numbers=false]{1.94\cdot 10^{11}}{\ampere\per\square\metre}$, and $v_{\mathrm{c}}=0.25$ to create the skyrmions corresponds to $j_c\approx \SI[parse-numbers=false]{1.94\cdot 10^{12}}{\ampere\per\square\metre}$.
And a dimensionless skyrmion speed of $v=0.2$ corresponds to the dimensionful skyrmion speed $v_{\mathrm{d}} = \frac{\gamma D}{M_{\mathrm{s}}} v \approx \SI[parse-numbers=false]{89.7}{\metre\per\second}$.

\section{References}

\providecommand{\newblock}{}

\end{document}